# THE HISTORY OF THE OBSERVATORY LIBRARY AT ØSTERVOLD IN COPENHAGEN, DENMARK

## Bertil F. Dorch
*University of Southern Denmark, Campusvej 55,*
*DK-5230 Odense, Denmark.*
E-mail: bfd@bib.sdu.dk

## and

## Jørgen Otzen Petersen
*Fuglsanggårds Allé 25, DK-2830 Virum, Denmark.*
E-mail: jotzenpetersen@gmail.com

**Abstract:** About fifty years after the work that astronomer Tycho Brahe carried out while living on the island of Hven had made him world famous, King Christian IV of Denmark built the Trinity Buildings in Copenhagen: a students' church, a university library, and an astronomical observatory at the top of the Round Tower. The Tower observatory was opened in 1642, and it housed the astronomers from the University of Copenhagen until 1861 when a new, modern observatory was built at Østervold in the eastern part of the city. In 1996, all the University astronomers from the observatories at Østervold and the small town of Brorfelde were relocated to the Rockefeller Buildings at Østerbro, and the two observatories were closed.

In this paper we focus on the library at the observatory in Østervold, and its subsequent fate following the close-down of that observatory.

## 1 INTRODUCTION

The Danish capital, Copenhagen, has a long history of astronomy, with King Christian IV authorizing the construction in 1642 of an astronomical observatory on the top of the Round Tower. In 1861 this was replaced by a new observatory at Østervold (in the eastern rampart of the city), away from city lights and noise, and this remained the home of astronomers from the University of Copenhagen until 1996.

The main purpose of this paper is to ensure that important knowledge about the Astronomy Library at the University of Copenhagen will not be forgotten. Some information about the library at the top of the Round Tower exists from around the year 1815, but knowledge about the 135-year period between 1861 and 1996, when the library at Østervold evolved from modest beginnings to become the official main astronomical library in Denmark, has not previously been published in English.[1]

In this paper we provide some background context on the history of astronomy in Denmark, before discussing the more recent history of astronomical libraries in Copenhagen and commenting on the future of astronomical collections in Denmark.

## 2 BACKGROUND HISTORY

Tycho Brahe, who was born as Tyge Ottosen Brahe (1546–1607), the son of Otto Brahe of Knudstrup Estate, discovered his New Star in the constellation of Cassiopeia on the evening of 11 November 1572. Brahe published the book *De Nova Stella in Anni 1572* in 1573, which brought him world fame. After this, King Frederik II gave Brahe unique opportunities: a grant of one to two percent of the State Budget for research in astronomy, which was then both the most useful branch among the natural sciences (for navigation and determining time) and the most interesting—for its view of the world. Besides that, working out horoscopes based on astrology was a valued task for astronomers at that time.

On the Danish island of Hven, which is now under Swedish rule, Brahe built Uraniborg around 1580 (Andersen, 2002: 98), a fine Renaissance castle, and later Stjerneborg (Star Castle), the most advanced observatory of its time. Until 1597, when Brahe's disagreements with King Christian IV forced him to leave the country, he and his associates improved the precision of stellar positions in the sky by a factor of about ten. Unfortunately, Uraniborg and Stjerneborg were both demolished soon after Tycho had left the island, and now there are only small remnants left. The testament of Tycho Brahe are his *De Nova Stella* and his star catalogue. There is no doubt that Tycho must have had a very well-stocked library for his time. He also had a paper mill and a printing press on Hven, so that he could publish his own books. Several of these works exist today at the Danish Royal Library in Copenhagen and the University Library of Southern Denmark in Odense, and are digitally available to the astronomical community (cf. Dorch et al., 2019).

Tycho Brahe's best and probably most famous apprentice, Christian Sørensen Longomontanus (1562–1647), became a Professor of Astronomy at the University of Copenhagen in 1621. He was the first Director of the Observatory in the Round Tower (or *Rundetårn*, in Danish), which was founded in 1637 and gradually began to function (Andersen, 2002: 101). The Tower was officially opened in 1642, the year mentioned in an inscription on it. This was only some sixty years after the building of Uraniborg and forty-five years after Tycho had to leave Hven. Fortunately, we still have King Christian's wonderful group of Trinity Buildings, including the Church of the Holy Trinity and an impressive library room, which housed the University Library until 1861, and where the astronomers worked with their telescopes on the top of the Round Tower.

Figure 1 shows the Trinity Complex, which today looks very similar to its original appearance in 1656, except for the small green dome at the top of the Round Tower. In the foreground of the photograph is one of the roofs of the Regensen (the Regent's Student College). This College, which was founded by Christian IV in 1623, has now supported close to 100 students over almost 400 years. One of the authors of this paper (JOP) lived there from 1957 to 1959.

The history of the Library from 1861 to 1996 is connected to the research and teaching of astronomy at the University of Copenhagen. In 1861 construction of the building at Østervold was finished (Figure 2), and the astronomers moved in. In January 1996 all the astronomers who were employed at the University of Copenhagen were moved from the observatories at Østervold and Brorfelde to the so-called Rockefeller Building, close to the National Hospital. For the locations of Copenhagen, Brorfelde and Hven see Figure 3.

## 3 THE YEARS 1861–1996: THE MAIN DEVELOPMENTS AT ØSTERVOLD

The use of the rooms in building no. 3 in Øster Voldgade street tells us that the astronomers have had varying conditions for working, and similarly for their library. In 1861, the Library was placed in quite a small room close to the main entrance. In 1961 the Library needed two rather large rooms on the ground floor plus areas in the basement. This shows that the Library was greatly enlarged when astronomy became more popular and especially when the number of books and journals grew. During the following years, this tendency continued, and in 1996, the Library took up almost twice as much space as in 1961. More shelves were placed in the largest of the Library rooms, and space in both the basement and against the wall in the circular aisle was now occupied by books.

In the annals of the University of Copenhagen, the history of the Library and the building itself can be followed. In the report for 1857–1863, the "Building of a new Obser-vatory" is described in detail (Anonymous, 1857–1863: 558–562). Money was set aside for the building itself and for acquiring two important items of equipment: a large refracting telescope and a meridian circle. The total cost of the new observatory was 91,463 Rigsdaler and 40 Skilling, of which 23,585 Rigsdaler and 85 Skilling were paid for the two main telescopes.

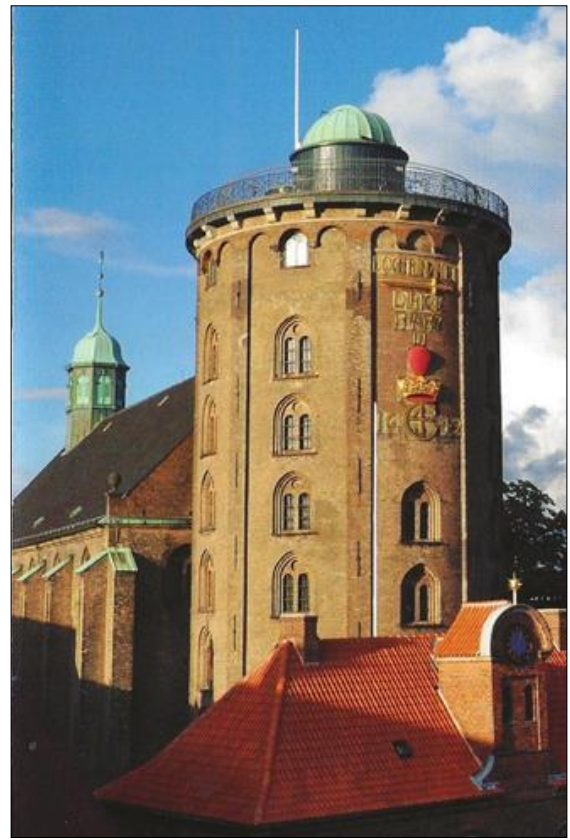

Figure 1: The Trinity Complex includes the Round Tower, the beautiful Trinity Church and the Library Hall situated just above the church itself and of precisely the same size as the church room. The gilt inscription at the top of the tower's façade is a rebus designed by Christian IV himself. It can be read as follows: Proper learning and justice (the golden sword) guide God (Jehova in Jewish letters) into the heart (red) of the crowned Christian IV, 1642 (photograph: brochure from the Observatory).

The report for 1920–1923, under the heading "The scientific Institutes of the University, The Astronomical Observatory", discloses that the annual expenditure granted by the exchequ-er to the Observatory (for operating the Observatory, but excluding salaries) was 7128 DDK (Danish kroner, in modern terms) (Anonymous, 1920–1923: 347). Of this total expenditure, 1300 DDK was spent on solid fuel. At this time, the building was heated by a variety of stoves, and there was an employee who took care of the heating arrangements. Besides the rooms for three telescopes on the ground floor and the big refracting telescope in the tower, the building contained four flats for the Professor, the Observer, the Scientific Assistant, and the Porter. In the budget, regarding a supplementary grant in 1923–1924, an amount of 6250 DDK was granted to cover a deficit in the accounts, justified by the fact that it had been found difficult to operate the Observatory properly on the existing grant; it had been nec-essary to discontinue some journal subscript-ions, and new journals had not been bound. It was pointed out to Professor Elis Strømgren (1870–1947; Figure 4), who was the Director of the Observatory from 1907 to 1940, that this supplementary grant should be taken as a reminder to plan his budgets in future so that deficits would be avoided.

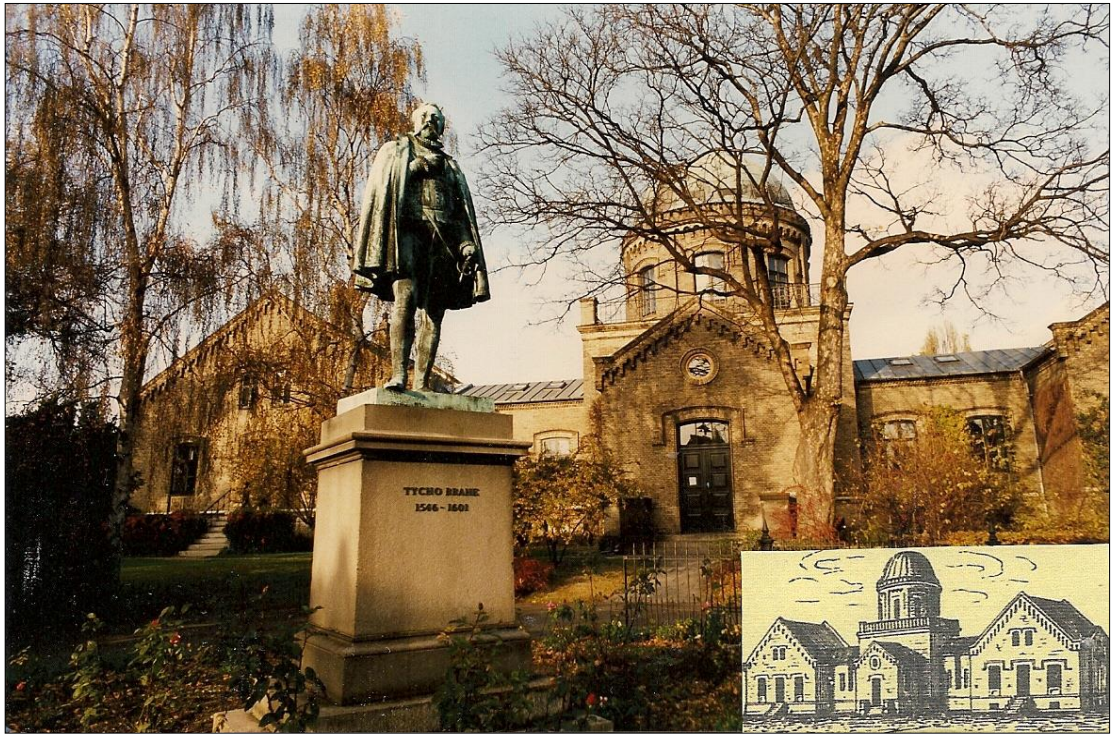

Figure 2: The University Observatory at Østervold, seen from the south, around 1961. The statue of Tycho Brahe is placed in a bed of roses. The bottom right-hand corner insert shows a sketch of the Observatory made in about 1861 that has been used on the cover of official Danish almanacs since 1864 (photograph: Observatory staff).

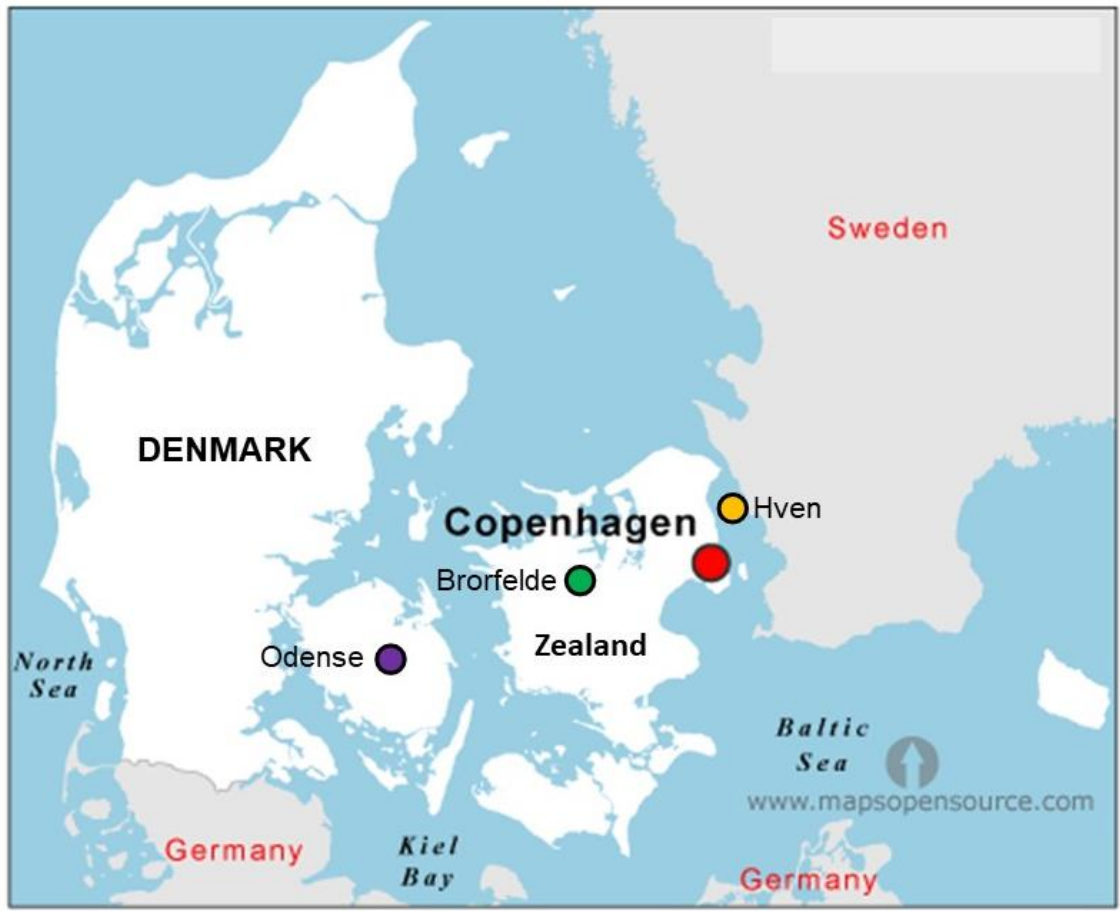

Figure 3: Map of Denmark and surrounding areas of Germany and Sweden, showing the island of Hven (now Swedish territory), and the the locations of Brorfelde Observatory and Copenhagen on the island of Zealand, and the University of Southern Denmark in Odense (www.mapsopensource.com; map modifictions Wayne Orchiston).

In the finance act of the annals of 1943–1944 we find a grant of 24,400 DDK for various construction works on the buildings, specifically for altering the three observing rooms in the middle of the building into a study and library rooms. However, because of the wartime scarcity of building materials this work was not carried out, but on 15 January 1944 the Ministry of Finance gave permission to spend 6000 DDK out of the grant on bricking up the observing slits in the walls and installing four windows in the building. The current harmonious façade facing south, and the less attractive northern one, were made in 1944, and the present arrangement of part of the ground floor (excluding the staff flats) is from the late 1940s. The bricked-up slits for observation are still clearly visible on the northern wall, while the southern façade was designed so carefully that the four new windows look exactly like those from 1861; but it is still possible to discern the observing slits.

The fireplaces used for heating were not replaced by central heating until the 1950s. The building shared its heating system with what was then the Chemical Laboratory and Museum of Mineralogy. Julie Vinter Hansen (1890–1960; Figure 5) has told one of us that the grant was put on the finance act several times, but nothing happened until she pointed it out. The work was eventually carried out.

Only a few astronomers were employed during the first one hundred years after 1861. The after 1861. The number of students of astronomy was just as modest. In 1957, there were two astronomers in permanent employment, and from five to seven students, two of whom were doing research work (for more details, see Petersen, 2015: 28). A few years earlier there were four astronomers working in Copenhagen, and three more permanent employees in the village of Brorfelde about 70 kilometers west of Copenhagen (see Figure 3). Between 1960 and 1975, about ten more astronomers were employed in permanent jobs in Copenhagen and Brorfelde, and about the same number in part-time jobs. Besides these, secretaries and technical workers were also hired. How was space found for so many people at the Observatory in Copenhagen?

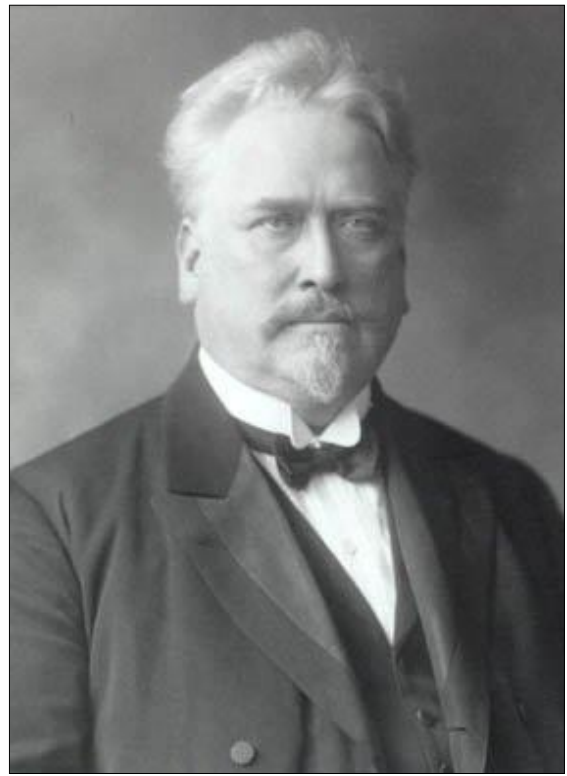

Figure 4: An undated photograph of Observatory Director Elis Strømgren (https://www.geni.com/people/Elis-Str%C3%B6mgren/6000000006970217298).

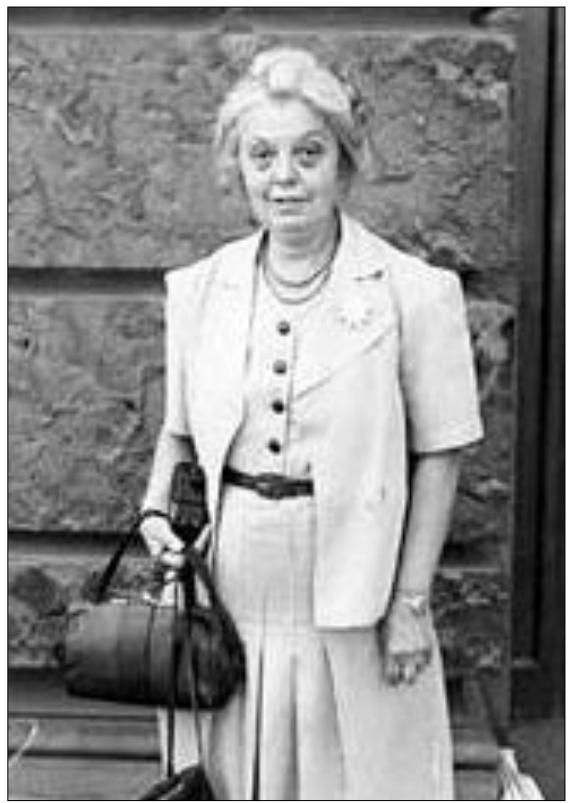

Figure 5: Julie Vinter Hansen, who was the Assistant, then the Astronomical Observer, and from 1951 to 1958 the Acting Director at the Observatory (https://en.wikipedia.org/wiki/Julie_Vinter_Hansen#/media/File:Julie_Vinter_Hansen.jpg).

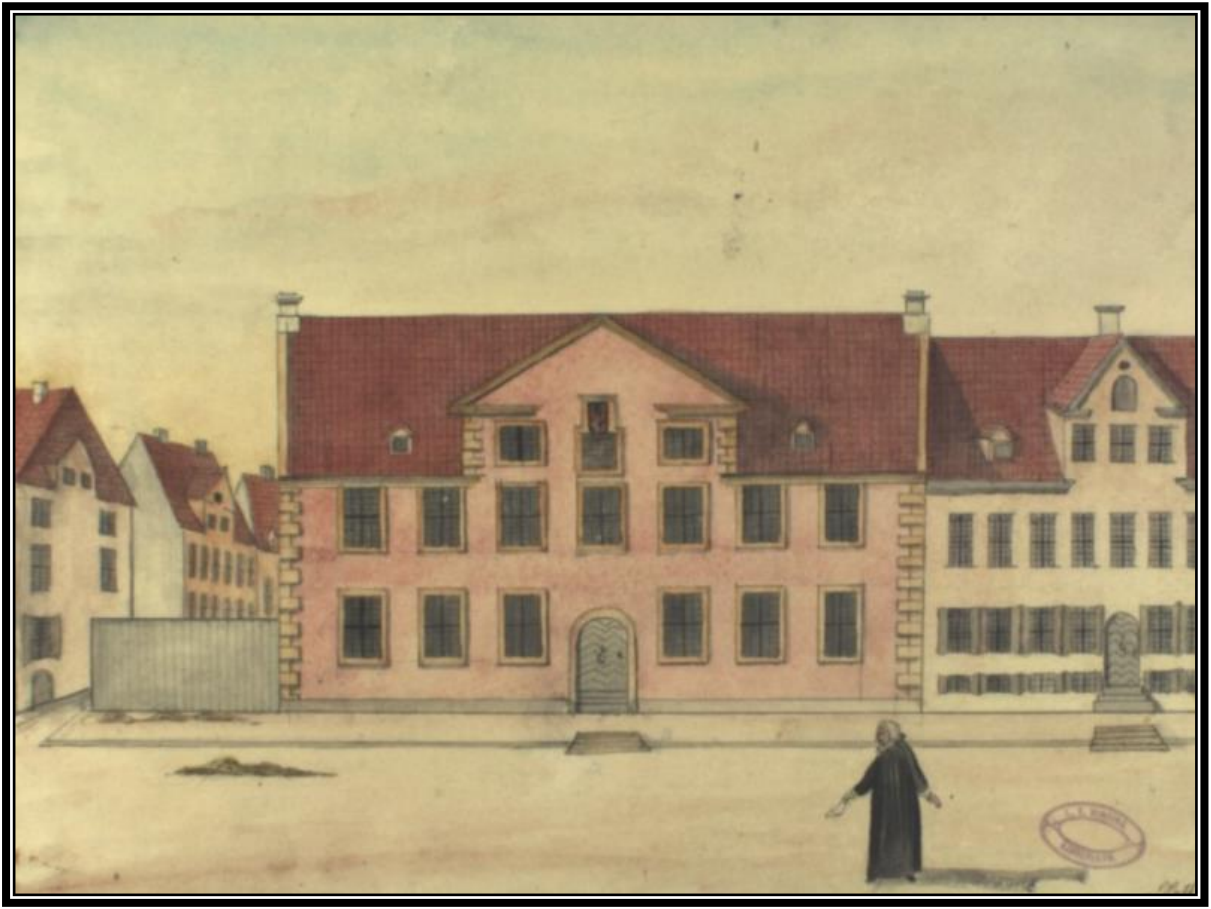

Figure 6: A painting of Valkendorfs Kollegium in 1749, after repairs following the Great Fire of Copenhagen in 1728 (https://en.wikipedia.org/wiki/Valkendorfs_Kollegium#/media/File:Valkendorfs_Kollegium_i_Sankt_Peders_Str%C3%A6de,_1749.png).

Prior to 1961, the entire east and west wings contained five large rooms (used as staff flats) and half a dozen smaller rooms. The flats gradually went out of use, and by 1975 the last astronomers had moved out. All the empty rooms could now be used as offices and rooms for computers and IT equipment. From 1958 the Astronomical Observatory had access to the first electronic computer in

Denmark, DASK, and in 1962, a GIER computer from the first series to be produced was placed in the building at Østervold. During the years 1975 to 1996, only small changes were made in the numbers of staff and students, since the grants were frozen during the challenging times of the 1980s. During this period, the main development was the increasing use of computers for astronomical calculations and office tasks.

## 4 ASTRONOMICAL LIBRARIES AND CATALOGUING

One of our earliest pieces of information about astronomy books at the University of Copenhagen is found in the history of the University, where we learn that the sixteen students at Valkendorfs College (Figure 6) had a small library of their own. The Danish statesman Christoffer Valkendorf (1525–1601; Figure 7) founded this College in 1589, and six years later a small private library was created comprising ninety-six volumes covering various subjects but concentrating mainly on mathematics, geography, and astronomy, with hardly any works on medicine or law (Ellehøj and

Grane, 1980: 306).

The first mention of an astronomical library in the Round Tower seems to be in an inventory list from 1815, comprising instruments and other furnishings that belonged to the Observatory. The list has been published in *Dansk Astronomi gennem 400 År* (Volume 3), and has 107 items (Gyldenkerne et al., 1990: 483–489). It was written and signed on 26 January 1815, by Mathias Bugge, who gave his guarantee that all the belongings of the Observatory were included in the list. Mathias Bugge was the son of Thomas Bugge (1740–1815), Professor and Director of the Observatory during the years 1777–1815. The list was made only eleven days after the death of Bugge the elder and must have been part of the transferral of the management of the institution. Items from 47 to 59 are books, astronomical ephemeral writings and tables (see Figure 8). This shows us that the astronomers had a small library of useful works. Some of them are also found in the library at Østervold, which must have inherited the 1815 collection directly. From the beginning in 1861, only very limited space had been set aside for an astronomical reference library. This was not very strange, since the permanent employees had their staff flats in the building; most of it was used for living in.

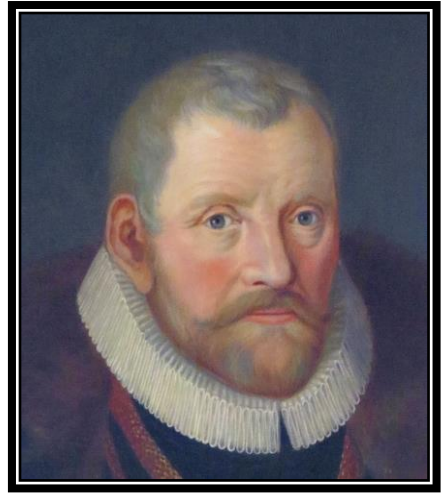

Figure 7: Christopher Valdendorf ca. 1600 (https://en.wikipedia.org/wiki/Christoffer_Valk endorff#/media/File:Christopher_Valkendorf_ til_Glorup.jpg).

The new, larger building for the University Library in Fiolstræde Street in Copenhagen was opened in 1861, as was the Observatory at Østervold. The next major expansion of the University Library took place in 1938, when

---

*V Bøger, astronomiske Ephemerider og Tabeller*

No.47.   40 Eksemplarer i raae Matrix af Th. Bugges: Observationes astronomicae in annis 1781, 1782 og 1783, institutæ, Havniæ 1784. Er af Forfatteren trykte paa egen Bekostning og skjænket til Observatoriet. Men Kobberpladerne mangler.

No.48.   Johannis Bayeri Uraniometria. Ulmæ 1661 in Fol.

No.49.   2de Protokoller ned rene Blade, hvori adskillige Standspersoner og Fremmede, som haver besøgt Observatoriet, har indskrevet deres Navne.

No.50.   Bodes mindre Himmelatlas, nyeste Udgave Berlin 1805 i Tværfolie og tilhørende Forklaring[?] med Titel: Forstellung der Gestirne, Berlin 1805, in 4to.

No.51.   Bodes allgm.Beschreibung d. Gestirne, nach Messergebnisse von 17240 Sternen, Berlin 1801, in Tot. (Denne bog er mig af min Fader afd. Etatsrad Bugge udlaant til brug paa Observatoriet, men den er i hans Regnskaber ingensteds ført Observatoriet til Udgift. Jeg er derfor i Tvivl om denne Bog maa betragtes som Observatoriets Ejendom, eller om den tilhører min afd. Faders Dødsbo).

No.52.   Bodes Verzeichniss von 5505 Sternen, nach Bobachtungen von Dr. Piazzi, Berlin 1805, in 4to.

No.53.   Samme Bog.

No.54.   De Zaith, Tabulae speciales aberrationes et nutationis Vol I og II. Gothae 1806, in 4to.

No.55.   Ejusdem, Tabulae motum Solis verae et iternem correctae. Gothae 1804, in 4to.

No.56.   Bodes astronomisches Jahrbuch for Aarene 1791 og 1792 samt 1798-1816. Dog mangler 1802. I alt 20 Vol. De øvrige Aargange ere brændte tilligemed min afd. Faders Proffessorresidents, under Bombardementet i 1807.

No.57.   Nautical=Almanac for Aarene 1798-1810, samt 1814 og 1815. Aargangen 1805 er in Duplo. I alt 16 Vol. De foregaaende Aargange brændte under ovennævnte Bombardement.

No.58.   Hells Ephemerides astronomicae for Aarene 1775, 78 og 98. De øvrige Aargange brændte under det ovennævnte Bombardement.

No.59.   Vega Logarithmisches=Trigonometrische Tafeln. 1'ste B. Leipzig 1797, 800.

Figure 8: An excerpt from Mathias Bugge's inventory list: numbers 47–59 (after Gyldenkerne et al., 1990(3): 487–488).

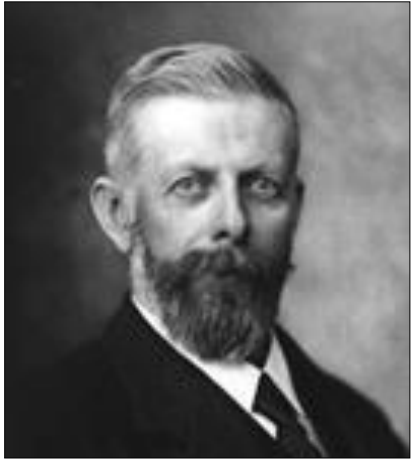

Figure 9: Mathematician Poul Heegaard (https://www.icmihistory.unito.it/portrait/heegaard.php).

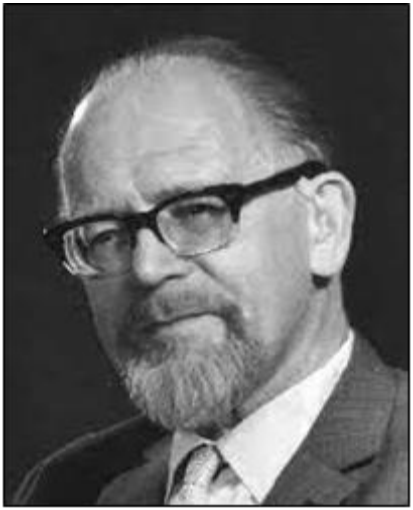

Figure 10: Statistician Georg Rasch (https://www.rasch.org/rasch.htm).

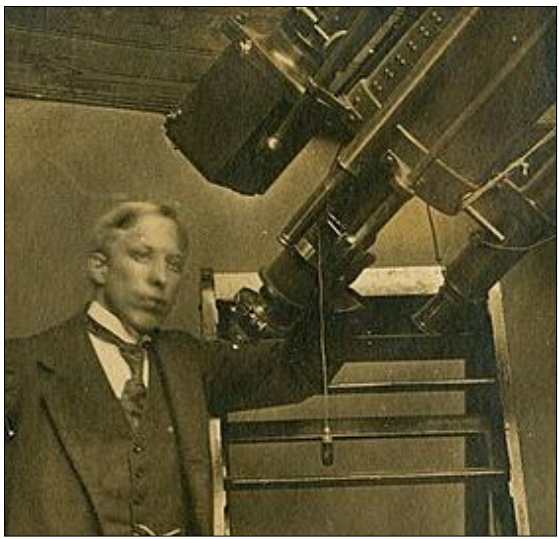

Figure 11: Danish astronomer Carl Luplau Janssen (https://commons.wikimedia.org/wiki/File:Carl_Luplau_Janssen_01.jpg).

works on medicine and natural sciences were moved from Fiolstræde Street to a new building in Nørre Allé street. This was known as the University Library's Second Department, and later the Danish Library for the Natural and and Medical Sciences. For a long time, the Østervold library was a part of this Library.

In 1990, Lisbeth Moustgaard, research librarian at the University Library, published her book *Uranias Tjenere: Episoder i Dansk Astronomi 1900−1950* (*Servants of Urania: Episodes from Danish Astronomy 1900−1950*) (Moustgaard, 1990), which tells about the institution, touching also on the Astronomical Observatory and its library. A marked, and now forgotten, feature in the first half of the twentieth century was the competition between Urania, the private observatory in Dronning Olgas Vej Street in Frederiksberg, and the Astronomisk Observatorium, the University Observatory at Østervold, involving both the library (especially for amateur astronomers) and scientific research and publishing, which will surprise most people today.

Circa 1900 the University Library was responsible for cataloguing the collections of the library at Østervold. In her book, Moustgaard (1990: 71−72) says:

> Around the turn of the century, the [University] Library had no employees with scientific or mathematical backgrounds. The first mathematician worked there in 1904-6, and in 1911 Luplau Janssen became the next employee with a degree in science working in the areas of mathematics and physics. Before this, in 1900-2, Poul Heegaard [1871−1948; Figure 9], a mathematician with special knowledge on astronomy, would assist the institution with tasks that required a knowledge of mathematics and astronomy. Among other things, he constructed the classification systems used for these subjects in the subject-specific catalogues of the Library. After a few decades, so great a development had taken place within both subjects that Heegaard's systems were no longer useful. Georg Rasch [1901−1980; Figure 10], a statistician, undertook the revision of Poul Heegaard's system for mathematics. Luplau Janssen [1889−1971; Figure 11] was responsible for constructing a new system for astronomy, and this was finished in the years 1937-39. (our English translation).

Lisbeth Moustgaard herself was in charge of the areas of mathematics and astronomy at the University Library for many years. It is likely that the system developed by Georg Rasch and Luplau Janssen was still in use when the file cards for the old Oservatory at Østervold were written and determined how the books were placed on the shelves (in 2018 at the Rockefeller Buildings). It is not yet clear how much of the catalogue from the file cards has been transferred into a database or how many of the books are accessible through the present library system.

## 5   THE FINE ØSTERVOLD BUILDING

As mentioned, in the 1850s the State granted money to build a new Observatory and purchase the two important new telescopes: a 26-cm refractor telescope, to be placed at the top of the building in a central dome, and a meridian circle (to continue the tradition established by Tycho Brahe and Ole Rømer).

The building was placed on the site of the Rosenborg fortification on the old ramparts at the edge of the new Botanical Gardens. Construction began in 1859 and was completed two years later. The architect of the Observatory was Christian Hansen (1803–1883; Figure 12), and during almost the same period he also designed the Municipal Hospital west of the Botanical Gardens. There are obvious similarities in the style of the two, although the Hospital is a much larger group of buildings. Hansen had visited Greece and became strongly interested in Byzantine buildings, particularly typical Byzantine churches, which have a square central room with a high dome and four transepts. This idea is found in both buildings and can be seen clearly in the plan of the Observatory (Figure 13), and perhaps less clearly in the case of the Hospital (see Figure 14).

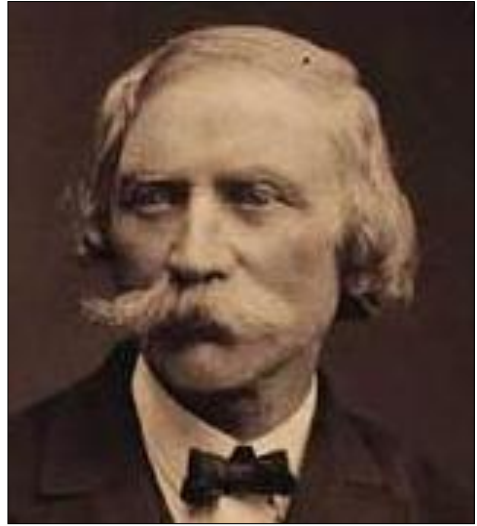

Figure 12: Danish architect Christian Hansen (https://upload.wikimedia.org/wikipedia/common s/7/71/Christian_Hansen.JPG).

Figure 13 shows how the rooms were divided in 1861. To the east and west of the ground floor telescope rooms led to the big staff flats for the resident astronomers: the Professor in the west wing, the Observer, Assistant, and Porter in the east wing. Apart from the influence from Byzantine churches, the building has reminiscences of the Round Tower, both in having telescopes situated at the top of a round tower, and in the building style itself, with pillars ending in brick half-moon arches.  There are also elements of temple architecture in the triangular gables towards south and north.

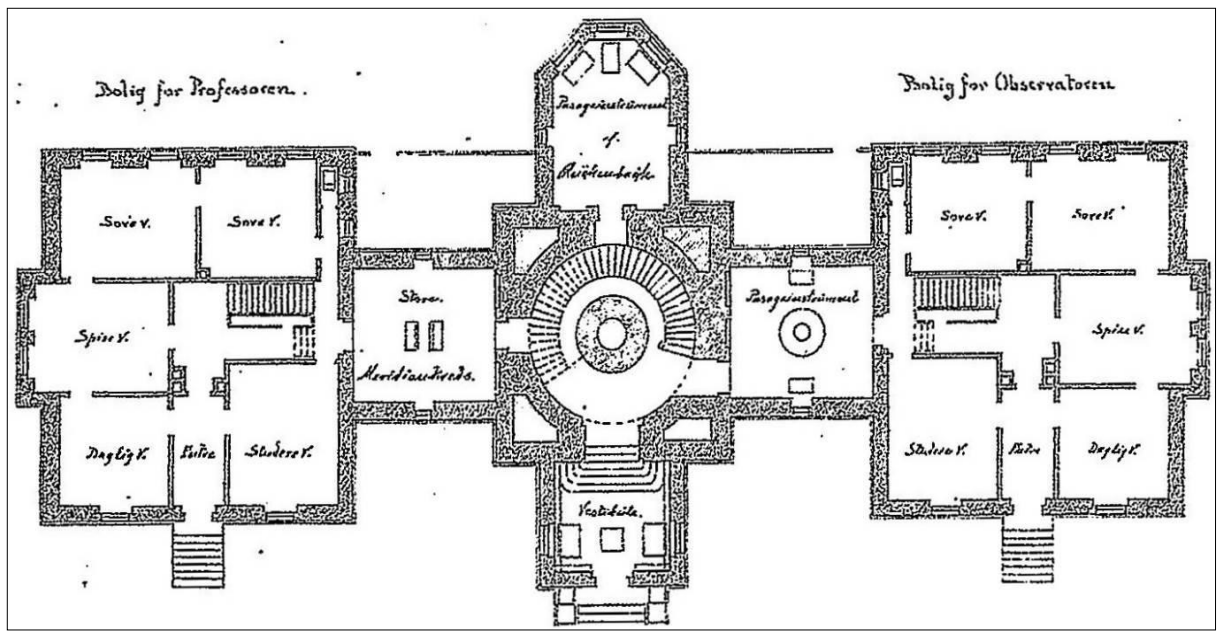

Figure 13: Floor plan for the ground floor of the Observatory in 1861. Like a Byzantine church, the Observatory had a central domed building and four transepts. The main entrance faces south. After passing through the small entrance hall, the visitor enters the circular aisle around the foundation of the big telescope, which was placed at the top of the central tower. On the ground floor, the two wings contained staff flats, the Professor living in the west wing and the Observer in the east wing. In 1957, the three telescope rooms, the study, and the basement in the observer's wing were converted for ordinary use, and after 1975 the entire area was used for offices and working rooms.

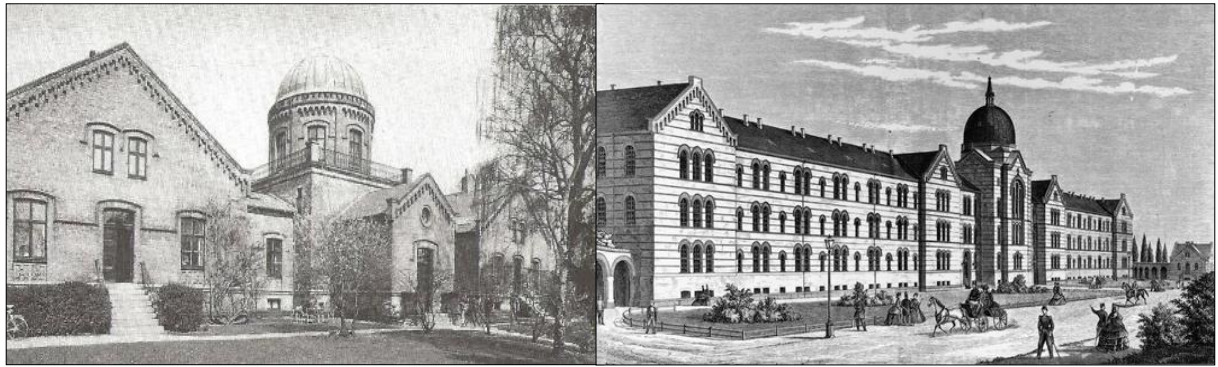

Figure 14:  On the left is the Observatory at Østervold photographed in 1986. The domed tower is in the centre, with transepts pointing to the four cardinal directions, exactly as in a church. This building does not have an open room beneath the dome, as the pier for the 26-cm refracting telescope goes from the dome-room down through the room below and into the ground.  From this ground floor room are steps on the southern side leading to a vestibule and the front door.  Stairs then lead down to the ground. The etching on the right shows the façade of the Municipal Hospital on Nørre Farimagsgade Street in 1863, soon after the building was constructed. Behind the big windows at the top of the central building was the church room beneath the dome. The frieze on the top had half-moon arches like those on the building at Østervold and on the Round Tower. The dome suggested the Byzantine church style (left photograph, courtesy: City Museum; etching on the right: https://www.wikiwand.com/en/Copenhagen_Municipal_Hospital).

Thanks to the floor plan from 1861, we know where the Observatory library was located.  At first, the idea was that the books were to be placed in the vestibule immediately by the front door.  It is obvious that any other location would be hard to find, since the rest of the large building was occupied by telescopes or staff flats.  It is also clear that this placement was an unfortunate one in case of rain or snow.  Some of the damage found on the old books today was probably caused by the damp conditions in the vestibule.  After about 1920, the smaller telescopes on the ground floor went out of use as noise and light from the city around the building were interfering with the observations.  In any case, by then pendulum clocks had become so precise that the time service did not need to rely on transit observations of stars anymore, and synchronization could be carried out using international radio signals.  As we have already noted, in the 1940s the three ground floor telescope rooms were altered and became work rooms, without slits in the roof for telescopes.  Meanwhile, windows were added, keeping to the original style of the wings.

When one of the authors of this paper (JOP) first saw the Observatory in 1957, both the big ground floor rooms labelled "Passageinstrument" and "Studere V." (Study Room) had been library rooms for a long time.  They took up most of the available space, since the large east and west wings only contained staff flats.  In the middle building, towards the west, there were two small offices for secretaries, and towards the north an auditorium, which was not in use for its original purpose, but was used for storage.  The study room in the Observer's wing was known as the Reading Room, and had filing cabinets and drawers with work materials, and bookshelves with reference books, journals and other reading materials.  It also had a handsome old clock.  In the middle of the room there were two shiny mahogany tables with comfortable chairs in mahogany and leather (Figure 15).  The furniture had probably been placed there by Julie Vinter Hansen, who had occupied the Observer's flat since 1923.  Besides these rooms, several rooms on the basement floor contained the collection of Observatory publications, older journals for which there was no longer room on the ground floor, and the Library archive, of such things as special prints, notebooks, maps, etc.  After Professor Bengt Strömgren had retired in 1987, books and papers from his office that seemed less interesting for inclusion in the library were placed here as well.  This included, for example, notebooks from his school days.  It is likely that all this was thrown out during or after the relocation to the Rockefeller Building in 1996.

## 6   JOURNALS, BOOKS, AND OBSERVATORY PUBLICATIONS

As in other libraries, cataloguing and accessibility determined the quality and usefulness of the Observatory Library.  The files at Østervold consisted of eighteen drawers of file cards, sized 12.4 × 7.4 cm.  Figure 15 shows the drawers against the west wall in the Reading Room;

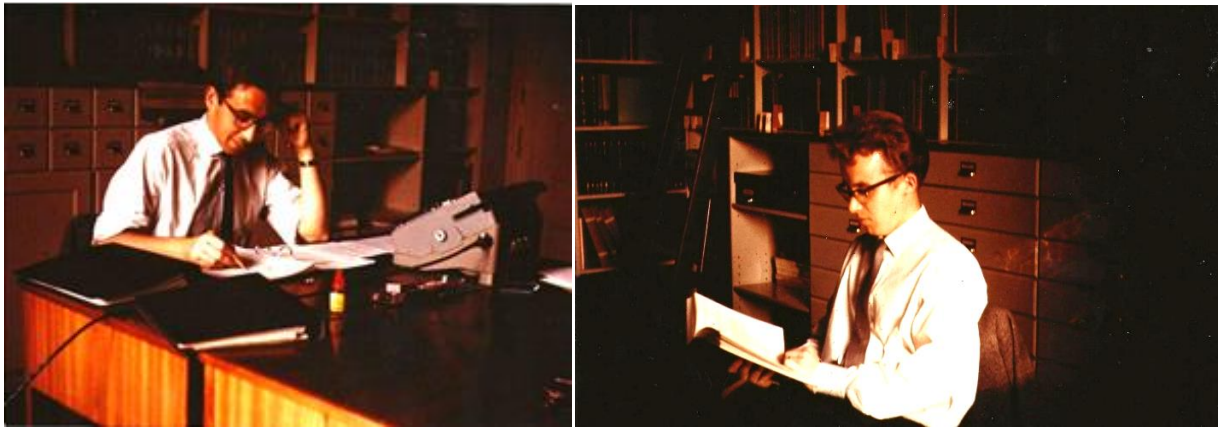

Figure 15: Two photographs of the Library Reading Room in 1960. On the left, a student, Henning Elo Jørgensen, who would later become the Professor of Astronomy and Head of Department, is using an electric calculating machine in front of the Library drawers. On the right, is Jørgen Otzen Petersen (the second author of this paper), a recent Master of Science, at the opposite working table.

the light comes from the south window. To the right of the files are two shelves, used by the librarians for mail etc., and the entrance door. Above the files, there are bookshelves. The shelves continued to the ceiling and contained as much of the collection of journals as there was room for. Since new journals were continually arriving, the oldest part of the collection had to be moved into the basement.

On the east wall, behind one of the authors (JOP), there are shelves and drawers, which are clearly seen in Figure 15. Around the year 1960, these drawers contained e.g. old lists and ledgers which were to be kept. There was also a drawer containing letters and works which were more or less unintelligible, mailed to the Observatory by more or less critical and/or hopeful authors with revolutionary new ideas (the flip drawer). They probably received a form letter in reply acknowledging receipt but stating that the Observatory regretted that staff were not in a position to comment.

Apart from journals, the space against the walls held a collection of the most important reference works for the use of employees and students. On some of the shelves there are small file card holders. They contain notes of loan, which had to be placed there instead of the book, so that it would always be clear where any individual book could be found.

The book collection was placed in the big room east of the circular aisle. The room had shelves from floor to ceiling along the walls, and bookshelves also took up a major part of the floor. At that time the books were arranged in groups according to a list that felt natural to astronomers:

A: Astronomy in general, biographies, bibliographies.
B: Practical and spherical astronomy, ob-servatory publications, star atlases, star charts, etc.
C: Celestial mechanics.
E: Mathematics and tables.
F: Astrophysics, related sciences (physics, optics, chemistry), space physics.
G: Journals and miscellaneous works.
S: Masters' Theses, Ph.D. dissertations.

At Østervold the books started by the circular aisle at the entrance, and the other groups followed, first along the walls, then on the bookshelves placed on the floor. In 2018 the books were still kept in this order in a basement room in the Rockefeller Building.

Publications from various observatories formed a special group. This collection had been carefully built over more than a century, and there are very few of its kind in the world. The background for the collection was a worldwide exchange of publications between observatories. For many years, perhaps since 1750, about 20 big, well-established observatories and various minor ones of varying quality had existed. Most observatories have taken pride in their employees' publications, cf. the important list of publications from all the institutes at the University of Copenhagen. Of course, many studies were published in major astronomy journals. But there were many that did not fit into any journals and could not be published in another established form such as a monograph. This kind of literature could be exchanged as special prints, which made them available to readers who, it was hoped, would be interested. It was prestigious for an observatory to exchange materials with many other observatories. The Observatory at Østervold spent many resources on special prints of published works and of local studies which were not published elsewhere for their list of exchange partners, and in return they received similar materials. One exotic address was the Purple Mountain Observatory, Nanjing, China; another was the Observatory of F.M. Bateson, Rarotonga, Cook Islands, South Pacific.

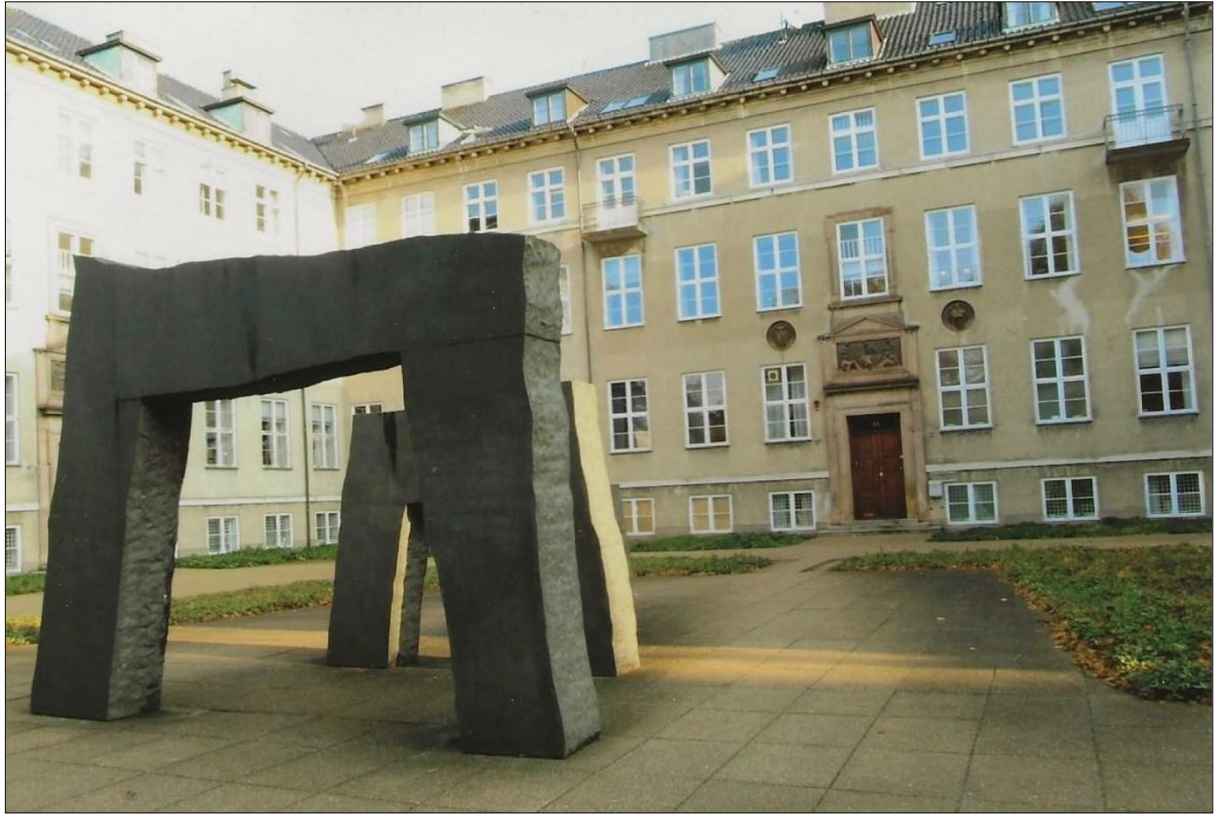

Figure 16: The yard of the three-winged Rockefeller Complex with the sculpture "Tre Porte" (Three Gates) by the artist Peter Bonnén around the year 2015. The sculpture leads to the main entrance. It is seen that the traditional, symmetrical layout of the building had four levels. Originally in 1928, the complex was donated by the Rockefeller Foundation in New York to Copenhagen University. Just before 1996 it was renovated to house astronomers, geophysicists and space physicists. Photo: Jørgen Otzen Petersen).

Obviously, the materials that the Library received over the years varied widely: from big heavy tomes to single pages of e.g. reports of a series of observations, and these materials turned up at very irregular intervals. The librarian had to collect everything from each observatory. This was done by using solid sheets of cardboard tied together with tape. Everything was carefully collected over the years, and it was important to have the materials bound when there was a certain amount of them, otherwise it would become almost impossible to keep track of them. Nevertheless, far from everything was bound. From Copenhagen, there is a bound series, *Publikationer og Mindre Meddelelser fra Københavns Observatorium* (*Publications and Minor Communications from the Copenhagen Observatory*), running from 1910 to 1970, with the numbers 1–213, which takes up some 90 cm. on the shelf. From no. 214, the series was renamed 'Copenhagen University Observatory Reprint' No. xyz. Each individual special print had this title stamped on it, and its number added by hand. 1995 was the last year in the series, containing the numbers 952–1026. Librarian Lone Gross told us that she used to write down a reference list of each number of the series annually. Unfortunately, only the 1995 list could still be found in 2019.

## 7 CONCLUDING REMARKS: THE FUTURE OF THE ASTRONOMICAL COLLECTIONS

As mentioned earlier, the employees of both the Østervold and the Brorfelde Observatories were transferred to the Rockefeller Buildings in 1996, on Juliane Maries Vej at the National Hospital (see Figure 16). Their libraries also were transferred. The book collection and the most important journals were placed in a big, airy room on the first floor. Since the collection from Brorfelde included works that the Østervold collection also owned, there were many duplicates, and the Library also acquired some minor collections of a miscellaneous character (mostly inherited from retired astronomers), it became necessary to use basement rooms as well to house the books.

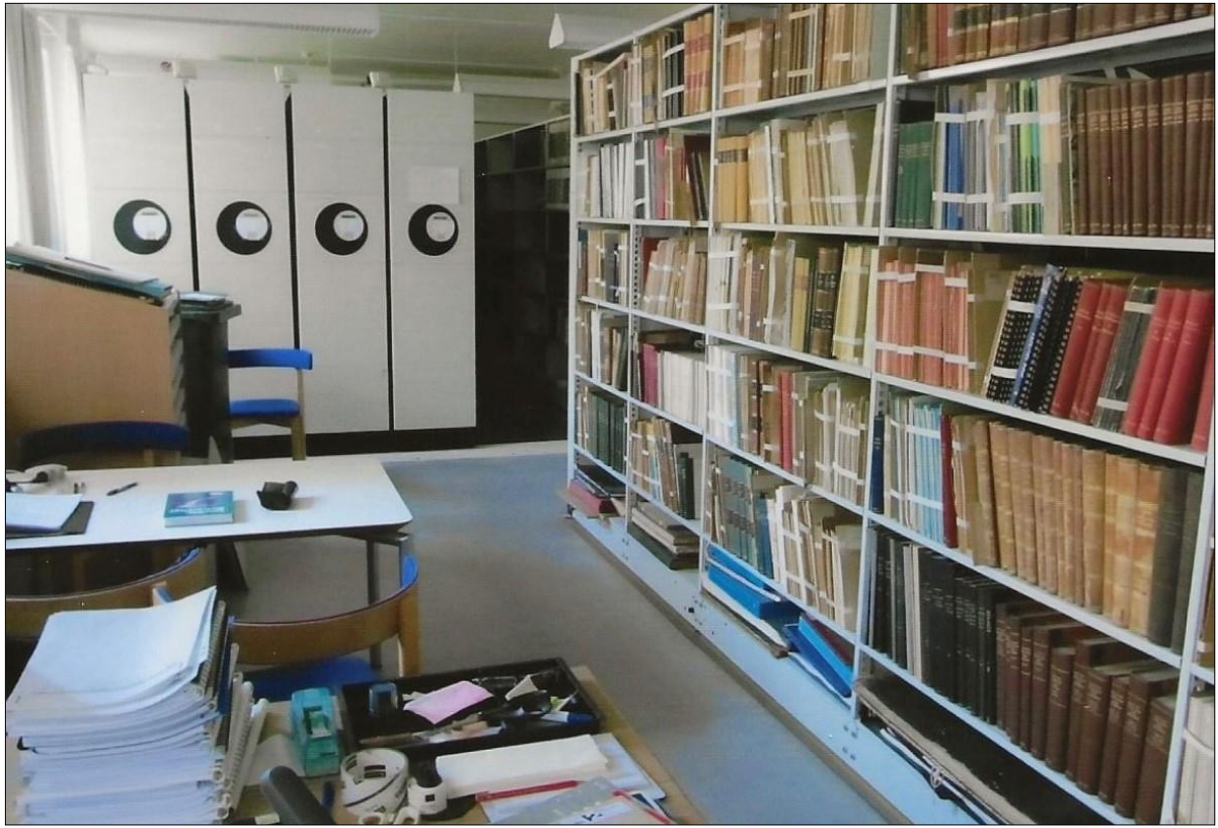

Figure 17: Details from one of the library rooms in 2018. All books are arranged in roller racks. In the foreground most of a roller rack containing observatory publications is seen. These collections contained about about 24,000 (photograph: Jørgen Otzen Petersen).

Around the year 2000, books from the Library were used less, since more and more information became easily available on the web. Nowadays, many researchers and students from the areas of physics and astronomy rarely use libraries at all. In the years 2000–2018 parts of the astronomy library were relocated a few times. In 2018 the entire book collection was placed in the basement, but was kept intact (see Figure 17). In 2015 it became clear that the Library would not be permanently kept in the Rockefeller Building, since the National Hospital was to take over the area for construction work. The description in the following is mainly based upon the personal experiences and judgments of the authors of this paper. The Rockefeller Building was demolished in 2021.

In 2015 Librarian Lisbeth Dilling made plans to revise and improve the collections on astronomy and geophysics in the Rockefeller Building. She expected, optimistically, that it would be possible to move the revised collection to the large new Niels Bohr Institute when the Library had to leave the Rockefeller Building. However, it became clear that the possibility of continuing all the activities at the Rockefeller Building would deteriorate, and that the Niels Bohr Institute would meet with problems because of delays during its construction. The situation grew still worse until 2018, when it seemed uncertain how the removal from the buildings was to turn out, and, especially for the Library, whether or not the book collection could be saved. It would, after all, be cheaper, easier, and save space, work, and time, to throw everything away. According to the rules for State property, a collection of this sort cannot be sold.

The University Library of Copenhagen and the Royal Library did not wish to receive the Østervold collection and keep it, so the Institute director, with help from the Library Board at the Niels Bohr Institute, had to look for other options at the eleventh hour, when it became clear that there was a risk that the collection would be thrown away.

Through an effort from especially Library Assistant Kader R. Ahmed and one of the authors of this paper (BD) as well as generous decisions from the Niels Bohr Institute and the faculty, almost all the important items in the Østervold Astronomy Library have been retained: the astronomy collection itself, books, collected works, calendars, star charts, star catalogues, etc., have now been mostly trans-ferred to the Physics Library at the Niels Bohr Institute at Blegdamsvej in Copenhagen, while the rare collection of observatory publications and many of the journals were taken over by the University of Southern Denmark in January 2019, where the collection is available for international historical and astronomical research (see Ellegaard and Dorch, 2021). This divided the former Østervold Library into two almost equal parts, with the works that are most likely to be requested by astronomers in Copenhagen, and the rare works of mainly historical interest placed in Odense.

## 6  NOTES

1. However, an account similar in part to what we present here was recently published by one of us in Danish in a publication on library history (Petersen, 2021).

## 7  ACKNOWLEDGEMENTS

The authors would like to thank librarians Lone Gross, Lisbeth Dilling, and Kader Rahman Ahmed warmly for their great help during many years, as well as for discussions and good advice in the preparation of the manuscript of this article. Thanks also to Steen Bille Larsen, historian and former Director of the National Library of Denmark, The Royal Library, for his careful reading of the Danish version of the manuscript.

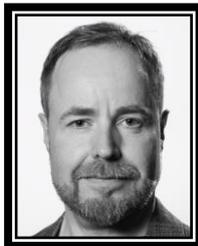

**Dr Bertil F. Dorch** was born in Denmark in 1971, and is an Associate Professor in the Department of Physics, Chemistry and Pharmacy at The University of Southern Denmark (SDU), and since 2013 Director of the Research and University Library at SDU. He received his PhD in Physics from The Niels Bohr Institute at the University of Copenhagen in 1998, where he also obtained his Master's, in Astronomy, in 1995. He previously held research positions at the Royal Swedish Academy of Science and at the University of Copenhagen.

Apart from teaching, Dorch participates in astronomy outreach, and is active in public and research political debate. Between 2014 and 2020 on three occasions he was President of the Danish Research Library Association. He is a member of the International Astronomical Union, and currently serves as a board member of various national and international organizations.

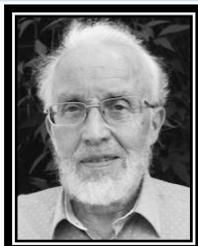

**Dr Jørgen Otzen Petersen** is born in Denmark in 1935, and since 1999 has been an Associate Professor Emeritus at the Niels Bohr Institute at University of Copenhagen. In 1960 he obtained an MSc in Astronomy from the University of Copenhagen, and in 1979 a doctorate, with a thesis on "Some problems in stellar evolution and stellar pulsation theories". Diagrams that display period ratios of double-mode radial pulsating stars are known as 'Petersen diagrams' after his work in the 1970s and 1980s.

As an Associate Professor, first at the Astronomical Observatory at the University of Copenhagen, and later at the Niels Bohr Institute, Petersen has acquired substantial first-hand knowledge regarding the routines and maintenance of the astronomical library, which spans nearly four decades.